\documentclass[10pt, conference, compsocconf]{IEEEtran}

\ifCLASSINFOpdf
   \usepackage[pdftex]{graphicx}
   \graphicspath{{figures/}}
 \else
\fi

\usepackage{array}
\usepackage{url}
\usepackage{subcaption}
\begin{document}
\title{Machine Learning DDoS
Detection for Consumer Internet of Things Devices}

\author{
\IEEEauthorblockN{Rohan Doshi}
\IEEEauthorblockA{Department of Computer Science\\
Princeton University\\
Princeton, New Jersey, USA\\
rkdoshi@princeton.edu}
\and
\IEEEauthorblockN{Noah Apthorpe}
\IEEEauthorblockA{Department of Computer Science\\
Princeton University\\
Princeton, New Jersey, USA\\
apthorpe@cs.princeton.edu}
\and
\IEEEauthorblockN{Nick Feamster}
\IEEEauthorblockA{Department of Computer Science\\
Princeton University\\
Princeton, New Jersey, USA\\
feamster@cs.princeton.edu}
}

\maketitle

\begin{abstract}
An increasing number of Internet of Things (IoT)
devices are connecting to the Internet, yet many of these devices are
fundamentally insecure, exposing the Internet to a variety of attacks. 
Botnets such as Mirai have used insecure consumer IoT
devices to conduct distributed denial of service (DDoS) attacks on critical
Internet infrastructure. 
This motivates the development of new techniques to automatically 
detect consumer IoT attack traffic. 
In this paper, we demonstrate that using
IoT-specific network behaviors
(e.g. limited number of endpoints and regular time intervals between
packets)
to inform feature
selection can result in high accuracy DDoS detection in IoT network traffic
with a variety of machine learning algorithms, including neural networks.
These results indicate that home gateway routers or other network
middleboxes could automatically detect local IoT device sources of DDoS
attacks using low-cost machine learning algorithms and traffic data that is
flow-based and protocol-agnostic.
\end{abstract}


\IEEEpeerreviewmaketitle

\section{Introduction}
The number of Internet of Things (IoT) devices is projected to grow from 8
billion in 2017 to 20 billion in 2020 \cite{McKinseyIoT}. Yet, many of these
IoT devices are fundamentally insecure.  One analysis of 10 currently popular
IoT devices found 250 vulnerabilities, including open telnet ports, outdated
Linux firmware, and unencrypted transmission of sensitive data~\cite{HP_IoT_Vulnerabilities,
BiTag}.

The proliferation of insecure IoT devices has resulted in a surge of IoT botnet attacks on Internet infrastructure.
In October 2016, the Mirai botnet commanded 100,000 IoT devices (primarily CCTV
cameras) to conduct a distributed denial of service (DDoS) attack against Dyn DNS
infrastructure~\cite{dyn-mirai}. 
Many popular websites, including Github, Amazon, Netflix, Twitter, CNN, and Paypal, were rendered inaccessible for several hours. 
In January 2017, the Mirai source code was publicly released;
DDoS attacks using Mirai-derived IoT botnets have since increased in frequency and
severity~\cite{Akamai_Mirai}.

This growing threat motivates the development of new techniques to identify
and block attack traffic from IoT botnets.  Recent anomaly detection research
has shown the promise of machine learning (ML) for identifying malicious
Internet traffic~\cite{AnomalyDetectionSurvey}.  Yet, little effort has been
made to engineer ML models with features specifically geared towards IoT
device networks or IoT attack traffic. Fortunately, however, IoT traffic is
often distinct from that of other Internet connected devices (e.g. laptops
and smart phones)~\cite{SmartHomeNoCastle}. For example, IoT devices often
communicate with a small finite set of endpoints rather than 
a large variety of web servers. IoT devices are also more likely to have repetitive
network traffic patterns, such as regular network pings with small packets at fixed time
intervals for logging purposes.

Building on this observation, we develop a machine
learning pipeline that performs data collection, feature extraction,  and
binary classification for IoT traffic DDoS detection. 
The features are designed to capitalize on IoT-specific network behaviors,  
while also leveraging network
flow characteristics such as packet length, inter-packet intervals, and protocol.
We compare a variety of classifiers for
attack detection, including random forests, K-nearest neighbors,
support vector machines, decision trees, and neural networks.

Given the lack of public datasets of consumer IoT attack traffic, we
generate classifier training data by simulating a
consumer IoT device network. We set up a local network comprised of a router,
some popular consumer IoT devices for benign traffic, and some adversarial devices performing
DoS attacks. Our classifiers successfully identify attack
traffic with an accuracy higher than 0.999. We found that random forest, 
K-nearest neighbors, and neural net classifiers were particularly effective. 
We expect that deep learning classifiers will continue to be effective with 
additional data from real-world deployments.

Our pipeline is designed to operate on network middleboxes (e.g. routers, firewalls, or network
switches) to identify anomalous traffic and corresponding devices that may be
part of an ongoing botnet. The pipeline is flow-based, stateless, and protocol-agnostic;
 therefore, it is well suited for deployment on consumer home gateway
routers or ISP-controlled switches. To our knowledge, this is the first
network anomaly detection framework to focus on IoT-specific features, as well
as the first to apply anomaly detection specifically to IoT botnets at the
local network level.
\section{Background and Related Work}
\label{section:related_work}

In this section, we present a brief background on network anomaly detection and middlebox limitations.

\subsection{Network Anomaly Detection}

Anomaly detection aims to identify patterns in data that do not conform to
expected behavior. In the context of our work, anomaly detection techniques
may be used to discern attack traffic from regular traffic. Simple
threshold-based techniques are prone to incorrectly classifying normal traffic
as anomalous traffic and are unable to adapt to the evolving nature of attacks
\cite{AnomalyDetectionSurvey}. More sophisticated anomaly detection
algorithms, particularly those using machine learning, can help minimize false
positives. Such approaches include deep neural networks, which promise to
outperform traditional machine learning techniques for sufficiently large datasets.

Anomaly detection has long been used in network intrusion detection systems
(NIDS) for detecting unwanted behavior in non-IoT networks.   The NIDS
literature can therefore inform the choice of anomaly detection methods for
IoT networks. In particular, the literature suggests nearest neighbor
classifiers~\cite{MINDS}, support vector machines~\cite{Eskin}, and rule-based
schemes like decision trees and random forests~\cite{QinAndHwang,ADAM}
as promising approaches.

Although there are parallels between NIDS and IoT botnet detection, there
has been little work tailoring anomaly detection specifically for IoT
networks. Our underlying hypothesis is that IoT traffic is different from
other types of network behavior. For example, while laptops and smart phones
access a large number of web endpoints due to web browsing activity, IoT
devices tend to send automated pings to a finite number of endpoints. IoT
devices also tend to have a fixed number of states, so their network
activity is more predictable and structured. For instance, a smart light bulb
could have three states: ``On," ``Off", and ``Connecting to Wi-Fi," each with
distinctive network traffic patterns. 

This hypothesis is
supported by the literature. Apthorpe et al. demonstrate how the finite states
of consumer IoT devices can actually be reflected in the repeated temporal
structures of send/receive traffic rates; this can even be used to infer
consumer usage behaviors \cite{SmartHomeNoCastle}. Similarly, the SCADA anomaly detection literature notes the unique network traffic patterns of sensors and controllers in infrastructure systems \cite{SCADA_AD1, SCADA_AD2}.  Miettinen et al.~further show how machine learning techniques can leverage the unique patterns of IoT network traffic for similar tasks, such as device identification \cite{SENTINEL}. Therefore, we use network traffic features that capture IoT-specific behaviors to better model IoT DoS attack traffic for anomaly detection.

\subsection{Network Middlebox Limitations}

Network middleboxes have limited memory and processing power, imposing
constraints on the algorithmic techniques used for anomaly detection. The
literature contains suggestions for how to meet these constraints. For
example, Sivanathan et al. investigated the use of software defined networks
to monitor network traffic at flow-level
granularity~\cite{FlowLevelNetworkAnalysis}. Their work suggests that using
flow-based features can be effective in detecting security threats without
incurring the high cost of deep-packet inspection. 
An anomaly detection framework for a consumer smart home gateway 
router should therefore have the following characteristics:
\begin{itemize}
\item{\textit{Lightweight Features}}.
Routers must handle high bandwidth traffic, so any features generated must be lightweight \cite{SNARE}. In order for an algorithm to scale to high bandwidth application, a given algorithm must rely on network flow statistics (how packets are sent) as opposed to deep packet inspection (what is in a packet).
\item{\textit{Protocol Agnostic Features}}. 
Routers must process packets from a variety of protocols (e.g. TCP, UDP, HTTP, etc.), so the algorithm must consider packet features shared by all protocols.
\item{\textit{Low Memory Implementation}}.
Routers are only able to maintain limited state due to memory constraints; caching adds latency and complexity. Thus, an optimal algorithm is either stateless or  requires storing flow information over short time windows only \cite{FlowLevelNetworkAnalysis}. 
\end{itemize}

\section{Threat Model}

\begin{figure}[t!]
\begin{minipage}{1\linewidth}
\begin{subfigure}[b]{\linewidth}
\includegraphics[width=\linewidth]{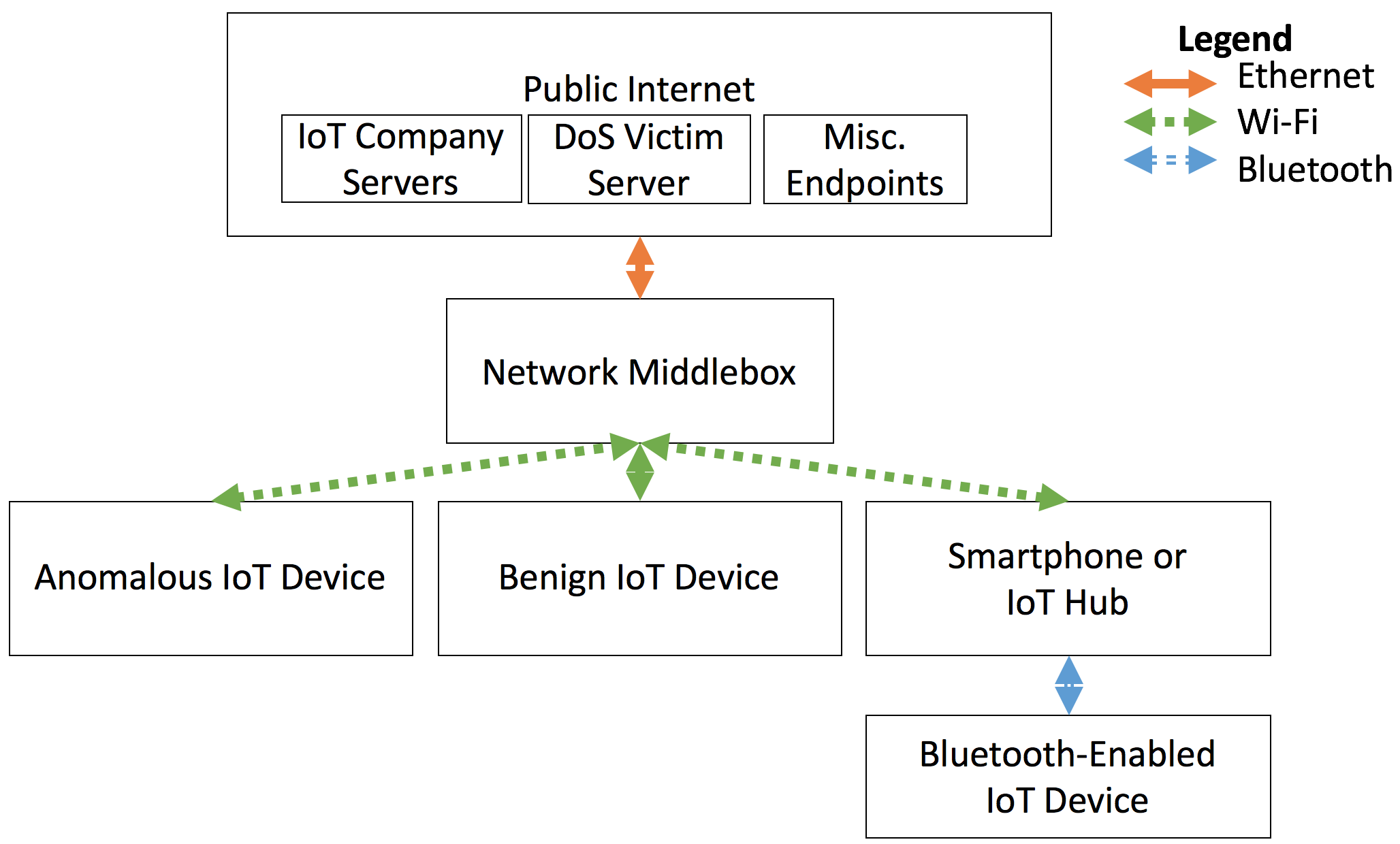}
\caption{Threat model\label{fig:threat_model_a}}
\vspace{12pt}
\end{subfigure} \hfill
\begin{subfigure}[b]{\linewidth}
\includegraphics[width=\linewidth]{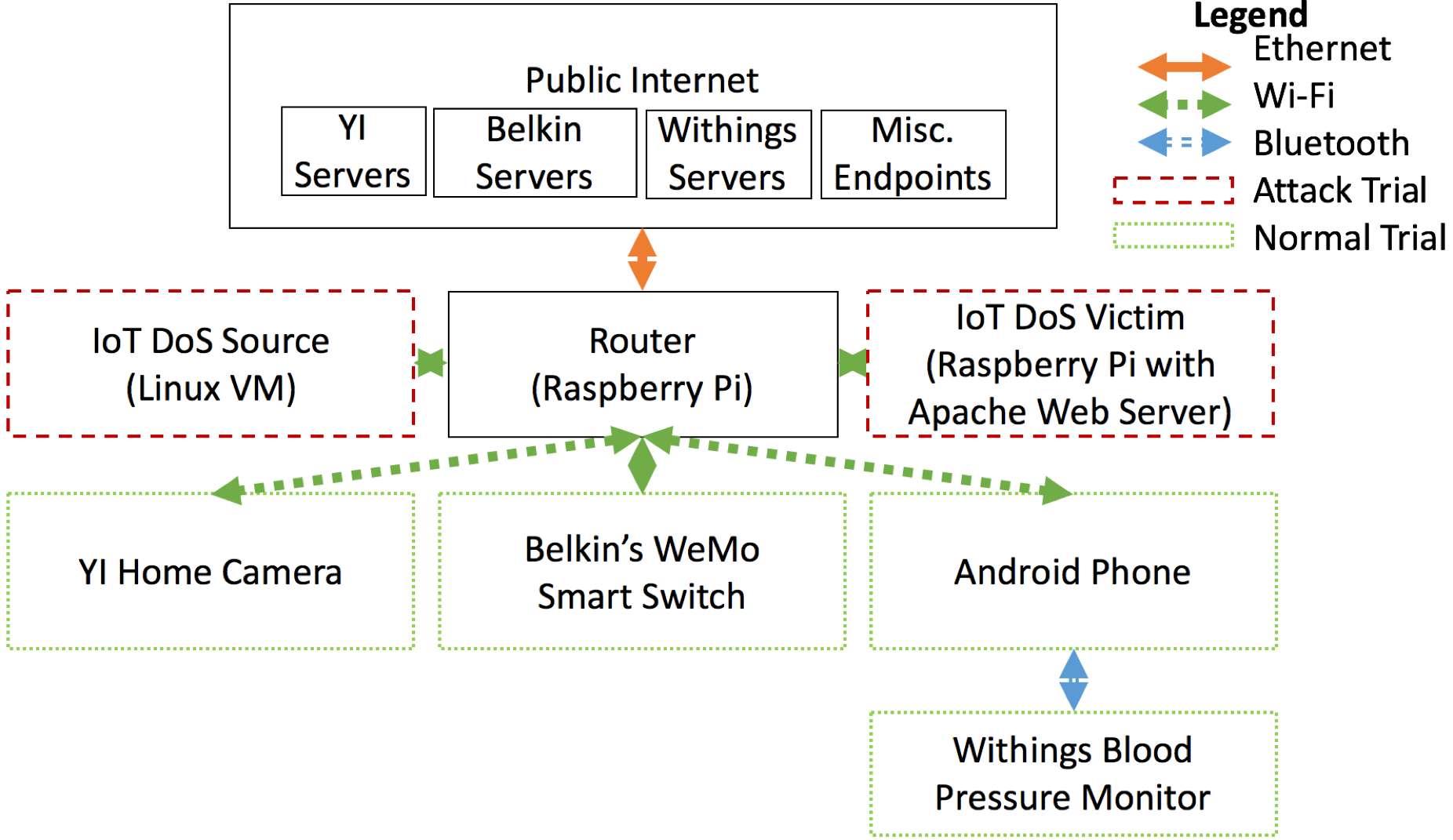}
\caption{Experiment setup\label{fig:experimental_setup}}
\end{subfigure}
\end{minipage}
\setlength{\belowcaptionskip}{-20pt}
\caption{Consumer IoT network threat model and corresponding experiment setup for collecting normal and DoS attack traffic training data.}
\label{fig:threat_model}
\end{figure}

Our threat model (Fig.~\ref{fig:threat_model_a}) makes various assumptions
about consumer IoT networks. We assume the network includes an on-path 
device, such as a home gateway router or other middlebox, that can
observe traffic between consumer IoT devices on the local network (e.g. a
smart home LAN) and the rest of the Internet. The device at this observation point
can inspect, store, manipulate, and block any network traffic that crosses its
path. All traffic between WiFi devices on the LAN or from devices to the
Internet traverses this middlebox.

Our goal is to detect and prevent DoS attack traffic originating from devices
within the smart home LAN. The DoS victim may be another device on the LAN
or elsewhere on the Internet. Any device connected to the
middlebox can send both network and attack traffic within the same time
period. Each device is also capable of conducting a variety of different DoS
attacks in series, and successive attacks can vary in duration. This reflects
how a remote botnet command and control (C\&C) may change orders. We
assume that the time range of DoS attacks are roughly 1.5 minutes, a
common duration for DoS attacks attempting to avoid detection
\cite{Akamai_Mirai}.

\section{Anomaly Detection Pipeline}
\label{section:fitbot_framework}

\begin{figure*}[tp]
\centering
\includegraphics[width=0.9\textwidth]{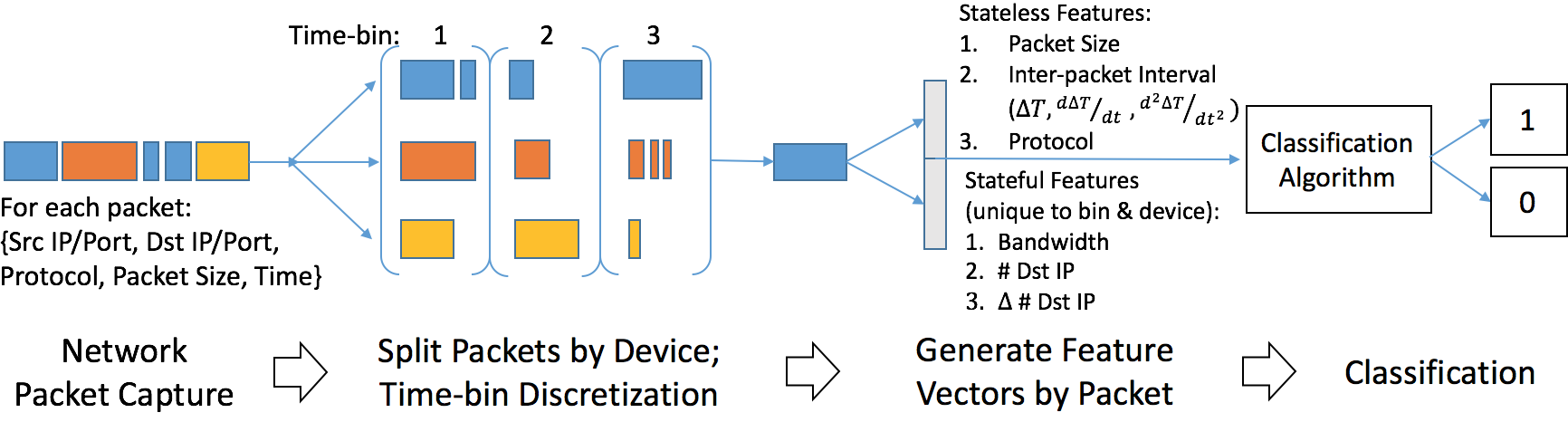}
\caption{IoT DDoS detection pipeline.}
\label{fig:fitbot_framework}
\end{figure*}

In this section, we present a machine learning DDoS detection
framework for IoT network traffic.  Our anomaly detection pipeline has four
steps (Fig.~\ref{fig:fitbot_framework}):

\begin{enumerate} 

\item {\em Traffic Capture.} The traffic capture process records the source
IP address, source port, destination IP address, destination port, packet
size, and timestamp of all IP packets sent from smart home devices.

\item {\em Grouping of Packets by Device and Time.} Packets from each IoT 
device are separated by source IP address. Packets from each device are further
divided into non-overlapping time windows by timestamps recorded at the
middlebox.

\item {\em Feature Extraction.} Stateless and stateful features are generated for
each packet based on domain knowledge of IoT device behavior. The stateless
features are predominantly packet header fields, while the stateful features
are aggregate flow information over very short time windows, requiring limited
memory to support on-router deployment.
(Section~\ref{section:feature_engineering}).

\item {\em Binary Classification.} K-nearest neighbors, random forests, decision trees, support vector
machines, and deep neural networks can differentiate normal traffic from DoS
attack traffic with high accuracy (Section~\ref{section:classification}).

\end{enumerate}
\subsection{Traffic Collection}

We set up a experimental consumer IoT device network to collect realistic
benign and malicious IoT device traffic~(Fig.~\ref{fig:experimental_setup}). We
configured a Raspberry Pi v3 as a WiFi access point 
to act as a middlebox.
We then connected a YI Home Camera~\cite{yi_camera} and Belkin WeMo Smart
Switch~\cite{wemo_switch} to the Raspberry Pi's WiFi network. A Withings Blood
Pressure Monitor was also connected by Bluetooth to an Android smartphone
associated with the WiFi network~\cite{withings}.

To collect normal (non-DoS) traffic, we interacted with all three IoT devices for
10 minutes and recorded {\tt pcap} files, logging all packets sent during that time
period.
We performed many interactions that would occur during regular device use, including
streaming video from the YI camera to the server in HD and RD modes,  turning the WeMo Smart Switch on/off and installing firmware updates, collecting blood pressure measurements from the Withing's Blood Pressure monitor, and sending the measurements to a cloud server for storage. 
We then filtered out all non-IoT traffic from the {\tt pcap} recordings, including background
traffic from the Android phone.

Collecting DoS traffic was more challenging. To avoid the security risks and complexity of running the real Mirai botnet code, we 
simulated the three most common classes of DoS attacks a Mirai-infected device will run: a TCP SYN flood, a UDP flood, and a HTTP GET flood~\cite{Akamai_Mirai}. 
We used a Kali Linux virtual machine running on a laptop as the DoS source, and a Raspberry Pi 2 running an Apache Web Server as the DoS victim. We connected both devices via WiFi to our Raspberry Pi 3 access point. The DoS source then targeted the victim's IP address with each class of DoS attack for approximately 1.5 minutes each. The access point recorded PCAPs of the attack traffic using the Linux dumpcap tool. The HTTP GET Flood was simulated using the Goldeneye tool \cite{Goldeneye}. The TCP SYN Flood and UDP Flood were simulated using Kali Linux's hping3 utility \cite{hping3}.
 
We then combined the DoS traffic with the normal traffic, spoofing source IP addresses, MAC addresses, and packet send times to make it appear as if the IoT devices simultaneously produced normal traffic and conducted DoS attacks. Each of the three IoT devices appeared to execute each of the three DoS attack classes once within a 10 minute internal. The attacks occurred in a random order for a random duration ranging uniformly from 90 to 110 seconds each. Thus, we collected roughly 300 seconds (5 minutes) of attack traffic per device. The distribution of attacks between devices was independent. 

This process produced a dataset of 491,855 packets, comprised of 459,565 malicious
packets and 32,290 benign packets. 
\subsection{Feature Engineering}
\label{section:feature_engineering}

\begin{figure*}[tp]
\centering

\begin{subfigure}[b]{0.25\textwidth}
        \centering
        \includegraphics[width=.90\linewidth, height=2.12in]{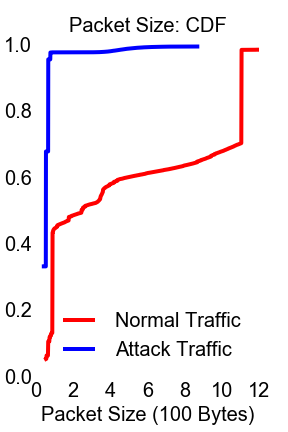}
        \vspace*{-2mm}
        \caption{}
        \label{fig:all-feature-cdfs-a}
\end{subfigure}%
\begin{subfigure}[b]{0.25\textwidth}
        \centering
        \includegraphics[width=.90\linewidth, height=2.12in]{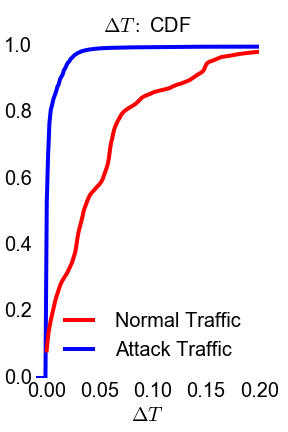}
        \vspace*{-2mm}
        \caption{}
        \label{fig:all-feature-cdfs-b}
\end{subfigure}%
\begin{subfigure}[b]{0.25\textwidth}
        \centering
        \includegraphics[width=.90\linewidth, height=2.12in]{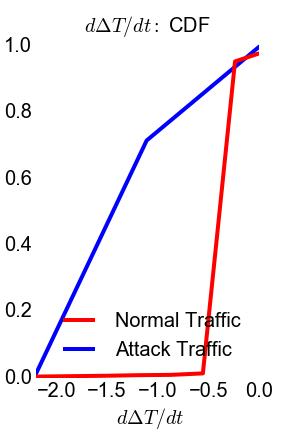}
        \vspace*{-2mm}
        \caption{}
        \label{fig:all-feature-cdfs-c}
\end{subfigure}%
\begin{subfigure}[b]{0.25\textwidth}
        \centering
        \includegraphics[width=.90\linewidth, height=2.12in]{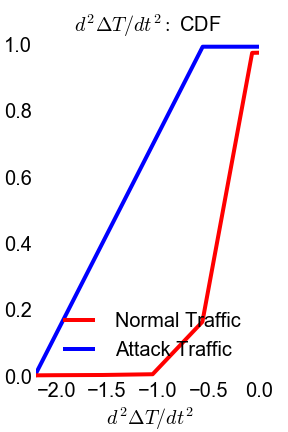}
        \vspace*{-2mm}
        \caption{}
        \label{fig:all-feature-cdfs-d}
\end{subfigure}

\vspace{1em}

\begin{subfigure}[b]{0.25\textwidth}
        \centering
        \raisebox{.16in}{\includegraphics[width=.90\linewidth]{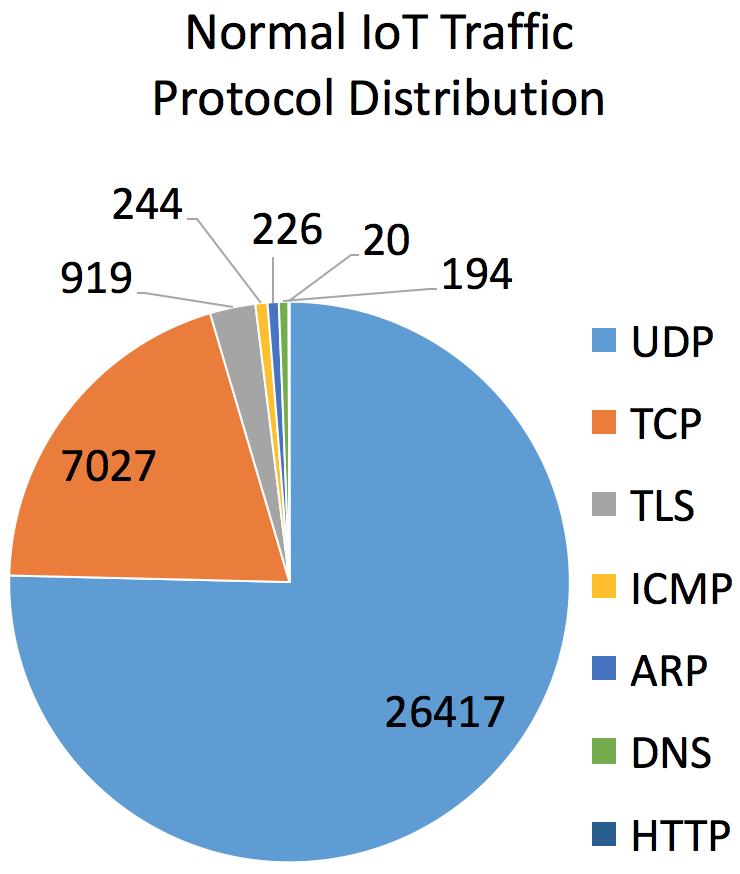}}
        \vspace*{-2mm}
        \caption{}
        \label{fig:all-feature-cdfs-e}
\end{subfigure}%
\begin{subfigure}[b]{0.25\textwidth}
        \centering
        \raisebox{.16in}{\includegraphics[width=.90\linewidth]{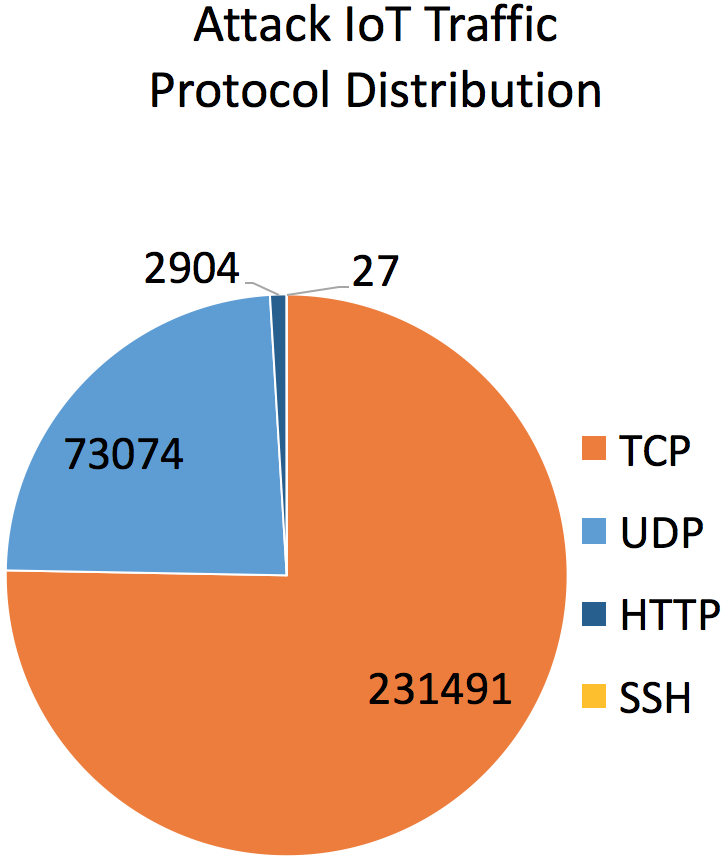}}
        \vspace*{-2mm}
        \caption{}
        \label{fig:all-feature-cdfs-f}
\end{subfigure}%
\begin{subfigure}[b]{0.25\textwidth}
        \centering
        \includegraphics[width=.90\linewidth, height=2.12in]{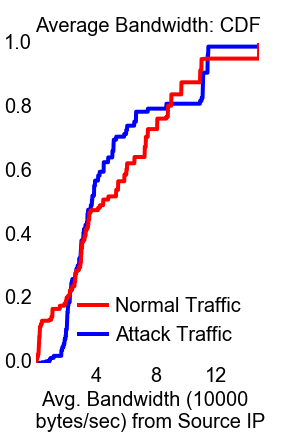}
        \vspace*{-2mm}
        \caption{}
        \label{fig:all-feature-cdfs-g}
\end{subfigure}%
\begin{subfigure}[b]{0.25\textwidth}
        \centering
        \includegraphics[width=.90\linewidth, height=2.12in]{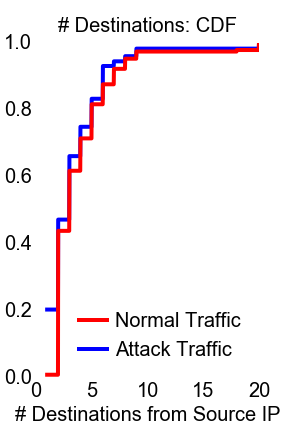}
        \vspace*{-2mm}
        \caption{}
        \label{fig:all-feature-cdfs-h}
\end{subfigure}

\caption{Comparison of feature statistics for normal versus DoS attack traffic. \textit{a)}~Packet sizes. \mbox{\textit{b--d)}}~Inter-packet intervals ($\Delta T$), $d \Delta T / {dt}$, and $d^2\Delta T / {dt^2}$. \mbox{\textit{e--f)}}~Protocol distributions. \textit{g)}~Average bandwidth over 10 second time windows. \textit{h)}~Number of unique IP destinations in 10 second time windows. }
\label{fig:all-feature-cdfs}
\vspace{-1.2em}
\end{figure*}

We explore two classes of features and analyze why they are
relevant to differentiating normal and attack IoT traffic.
{\em Stateless features} can be derived from
flow-independent characteristics of individual packets. These features are
generated without splitting the incoming traffic stream by IP source. Thus,
these features are the most lightweight.  {\em Stateful features}
capture how network traffic evolves over time. There is inherent overhead in
generating these features, as we split the network traffic into streams by
device and divide the per-device streams into time windows. The time
windows serve as a simple time-series representation of the devices' evolving
network behavior. These features require aggregating statistics over multiple
packets in a time window; the middlebox performing
classification must retain state, but the amount of state can be limited by using
short (e.g. 10-second) time windows.

\vspace{12pt}

\subsubsection{Stateless Features}

\paragraph{\bf Packet Size} The distribution of packet sizes differs significantly between attack and normal traffic
(Fig.~\ref{fig:all-feature-cdfs-a}). Over
90\% of attack packets are under 100~bytes, while normal packets vary between
100 and 1,200 bytes. 
A device conducting
a DoS attack, such as a TCP SYN Flood, is trying to open as many connection
request as possible with the victim to exhaust the victim server's resources.
Thus, an attacker wants to keep the size of the packets as small as
possible in order to maximize
the number of connection requests per second. In comparison, normal traffic can
range from simple server pings indicating that the device is
active (small packets) to video streaming data (large packets).

\paragraph {\bf Inter-packet Interval} Normal IoT traffic has limited
burstiness (Fig.~\ref{fig:all-feature-cdfs-b}-d). Most packets are sent at
regular intervals with appreciable time between packets. This may reflect IoT
network pings or other automated network activities. In contrast, a vast
majority of DoS attack traffic has close to zero inter-packet intervals ($\Delta T$)
and high first and second derivatives of inter-packet intervals. Using $\Delta T$, $\frac{d\Delta T}{dt}$,
and $\frac{d^2\Delta T}{dt^2}$ as features allows a classifier to capitalize on
this difference between normal and DoS traffic.

\paragraph{\bf Protocol}
Normal and DoS attack traffic also have varying protocol distributions (Fig.~\ref{fig:all-feature-cdfs-e}-f).
UDP packets outnumber TCP packets in normal traffic by almost a factor of three due to UDP video streaming. In comparison, TCP packets outnumber UDP packets in attack trafic by almost the same ratio. 
Attack traffic also includes fewer protocols in total. 
We capture protocol differences in a feature with a 
one-hot encoding of the three most popular attack protocols (IS\_TCP, IS\_UDP, and IS\_HTTP) and another binary indicator to reflect all other types of protocols (IS\_OTHER).  This captures the most popular protocols while minimizing noise and unnecessary dimensionality associated with less relevant protocols.

\vspace{7pt}

\subsubsection{Stateful Features}

\paragraph{\bf Bandwidth} The literature contains evidence that bandwidth usage
can be used to characterize network traffic patterns of IoT devices. For example, Apthorpe et al. were able to characterize consumer IoT device usage patterns from send/receive rates, but dividing network traffic by source device was necessary for the analysis \cite{SmartHomeNoCastle}. Similarly, our pipeline splits network traffic by source device and calculates the average bandwidth within 10-second time windows to measure the instantaneous bandwidth associated with each device. There are minor distributional differences in bandwidth usage between the normal and attack traffic (Fig.~\ref{fig:all-feature-cdfs-g}). We predict that a ML model will be able to leverage these differences.

\paragraph{\bf IP Destination Address Cardinality and Novelty} 
IoT devices are characterized by the limited number of endpoints with which they communicate \cite{SmartHomeNoCastle}. For example, a WeMo smart switch communicates with only four endpoints for the purposes of activation/deactivation from the cloud, retrieving firmware updates, and logging its status. Another key characteristic of IoT device traffic is that the set of destination IP addresses rarely changes over time.

We craft two features to reflect this behavior. First, a count of distinct destination IP addresses within a 10-second window; more endpoints may indicate attack traffic. Second, we calculate the change in the number of distinct destination IP addresses between time windows; new endpoints might suggest that the device is conducting an attack. 
Fig.~\ref{fig:all-feature-cdfs-h} supports the importance of these two features. Packets associated with attack traffic are in contact with, on average, more endpoints. This minor distributional difference can be leveraged in differentiating normal and attack traffic.

\section{Results}
\subsection{Classification}
\label{section:classification}

We tested five machine learning algorithms to distinguish normal IoT packets from DoS attack packets:
\begin{enumerate}
\item  K-nearest neighbors ``KDTree" algorithm (KN) 
\item Support vector machine with linear kernel  (LSVM) 
\item Decision tree using Gini impurity scores  (DT) 
\item Random Forest  using Gini impurity scores (RF)
\item Neural Network (NN): 4-layer fully-connected feedforward neural network (11 neurons per layer), trained for 100 epochs with batch size 32 using binary cross-entropy loss; hyperpameters chosen by optimization on a validation set 
\end{enumerate}
\noindent
We implemented these machine learning models using the Scikit-learn
Python library~\cite{SciKit}, except for the neural network, which was implemented
using the Keras library~\cite{chollet2015keras}.  All hyper-parameters were the
default values unless otherwise noted.

We trained the classifier on a training set with 85\% of the combined normal and DoS traffic and calculated classification accuracy on a test set of the remaining traffic (Table~\ref{table:results}). 
The accuracies of our four classifiers ranged from approximately $0.91$ to $0.99$.
Note that there are almost 15 times as many attack packets as there are normal packets due to the flooding nature of the DoS attacks. Thus, a naive baseline prediction algorithm that predicts that all packets are malicious would achieve a baseline accuracy of~$0.93$.

\begin{table}[tp]
  \caption{IoT Traffic Classification Results}
  \centering
  \begin{tabular}{llllll} \hline
      & \textbf{KN} & \textbf{LSVM} & \textbf{DT} & \textbf{RF} & \textbf{NN} \\\hline
    Precision (Normal)   &  .998 & .992 & .996  &  .999 & .983 \\ 
    Precision (Attack)   &  .999 & .991 & .999  &  .999 & .999 \\ 
    Recall (Normal)      &  .993 & .870 & .993  &  .998 & .989 \\
    Recall (Attack)      &  .999 & .999 & .999  &  .999 & .998 \\ 
    F1 (Normal)          &  .995 & .927 & .994  &  .998 & .986 \\ 
    F1 (Attack)          &  .999 & .995 & .996  &  .999 & .999 \\
    Accuracy             &  .999 & .991 & .999  &  .999 & .999 \\ \hline
  \end{tabular}
  \label{table:results}
\end{table}

The linear SVM classifier performed the worst, suggesting that the data is not linearly
separable.
The decision tree classifier performed well, achieving an accuracy of $0.99$, suggesting that the data can be segmented in a higher feature space.
The K-nearest neighbors classifier also achieved the same accuracy, suggesting that the two different data classes clustered well in feature-space. 
The neural network performed surprisingly well despite having fewer than half a million training samples from a 10-minute packet capture. 
Given the nature of the algorithm, the neural network is expected to scale its performance with the amount of available training data. 

\subsection{Feature Importance}

\begin{table}[tp]
  \caption{Feature Importance using Gini Impurity Scores.}
  \centering
  \begin{tabular}{ll} \hline
    \textbf{Feature} & \textbf{Gini Score} \\ \hline
    Packet Size            &  .510  \\  
    is\_HTTP               &  .177  \\ 
    $\Delta{T}$              &  .070  \\ 
    is\_TCP                &  .068  \\ 
    is\_OTHER              &  .043  \\ 
    is\_UDP                &  .041  \\ 
    $d\Delta{T}/dt$          &  .018  \\ 
    $d^2\Delta{T}/dt^2$      &  .012  \\ 
    Bandwidth              &  .006  \\ 
    \# Destinations                 &  .004  \\ 
    $\Delta$ \# Destinations          &  .003  \\ \hline
  \end{tabular}
  \vspace{10pt}
  \label{table:feature_importance}
\end{table}

The stateless features greatly outperformed the stateful features, as indicated by Gini impurity score (Table~\ref{table:feature_importance}). 
We expected this result, since the differences in the cumulative distributions of
normal and attack traffic were more pronounced than those of the stateless features (Fig.~\ref{fig:all-feature-cdfs}).
This result suggests that real-time anomaly detection of IoT attack traffic
may be practical because the stateless features are
lightweight and derived from network-flow attributes (e.g. 5-tuple and packet
size).

Incorporating stateful features nonetheless improved accuracy compared to
classification with the stateless features alone
(Table~\ref{tbl:IoT_feature_importance}). All of the classifiers experienced
 a 0.01 to 0.05 increase in F1 score by including stateful features.
This demonstrates that applying domain knowledge about IoT device
behaviors to feature engineering can enhance DoS
detection performance.

\begin{table}[tp]
  \caption{Classifier performance, with and without IoT-specific stateful (temporal) features.}
  \centering
  \begin{tabular}{p{2.2cm}lllll} \hline
     \raggedright \textbf{F1 (Normal)} & \textbf{KN} & \textbf{LSVM} & \textbf{DT} & \textbf{RF} & \textbf{NN} \\\hline
      \raggedright Stateless Features &  .967 & .920 & .977  & .981 & .939 \\
    \raggedright All Features & .995 & .921 & .995  &  .998 & .989 \\ \hline
  \end{tabular}
  \label{tbl:IoT_feature_importance}
\end{table}
\section{Discussion \& Future Work}

This preliminary work demonstrates that simple classification algorithms
and low-dimensional features can effectively distinguish normal IoT device traffic
from DoS attack traffic.  This result motivates follow-up research to evaluate
IoT DoS detection in more real-world settings.

First, we would like to replicate the results of this study with normal
traffic from additional IoT devices and with attack traffic recorded from a
real DDoS attack. This could involve using published code to create an IoT
device botnet on a protected laboratory network or collaborating with an ISP
to obtain NetFlow records or packet captures recorded during a DDoS attack.
This will be an essential test of the method's external validity.

Collecting a larger dataset would also allow us to see how DoS detection
accuracy is affected by the amount and diversity of IoT traffic. The network
behavior of IoT devices varies widely by device type \cite{SmartHomeNoCastle}.
We are curious whether certain types of devices are more amenable to network
anomaly detection, perhaps because their normal traffic follows more regular
patterns, or vice versa.

We would also like to experiment with additional features and more complex
machine learning techniques beyond those discussed in this paper. We believe
that there is great potential for the application of deep learning to anomaly
detection in IoT networks, especially for detecting attacks that are more
subtle than DoS floods. We hope that this work inspires further efforts to
develop network protection techniques specifically designed for IoT devices.

It is also an open question how best to intervene once an IoT device is
discovered to be part of a DDoS attack. Simply cutting the device off from
the network might not be feasible, especially if the device is essential
(e.g. a blood sugar monitor or a home water pump), because many smart devices
do not retain basic functionality without network connectivity~\cite{apthorpe2017closing}.
Notifying the user is an option, but many users of
home IoT devices will be unequipped to perform device maintenance beyond
powering off or disconnecting the device. 
\section{Conclusion}
\label{section:conclusion}
In this work, we showed that packet-level machine learning DoS detection can
accurately distinguish normal and DoS attack traffic from consumer IoT
devices.  We used a limited feature set to restrict computational overhead,
important for real-time classifcation and middlebox deployment. Our choice of
features was based on the hypothesis that network traffic patterns from
consumer IoT devices differ from those of well-studied non-IoT networked
devices.
We tested five different ML classifiers on a dataset of normal and DoS attack traffic collected from an experimental consumer IoT device network. All five algorithms had a test set accuracy higher than $0.99$.
These preliminary results motivate additional research into machine learning
anomaly detection to protect networks from insecure IoT devices.

\section*{Acknowledgments}

We thank Dillon Reisman, Daniel Wood, and Gudrun Jonsdottir. This work was
supported by the Department of Defense through the National Defense Science
and Engineering Graduate Fellowship (NDSEG) Program, a Google Faculty Research
Award, the National Science Foundation, and the Princeton University
Center for Information Technology Policy Internet of Things Consortium.

\bibliographystyle{IEEEtran}
\bibliography{IEEEabrv,MachineLearningDDoS}

\end{document}